\documentclass[final,5p,times,twocolumn]{elsarticle}

\usepackage{amssymb}
\usepackage{amsmath}
\usepackage{booktabs}
\usepackage{multirow}
\usepackage[T1]{fontenc}
\usepackage[colorlinks, citecolor=blue]{hyperref}
\usepackage{ulem}

\journal{Pattern Recognition Letters}

\begin{document}

\begin{frontmatter}



\title{MAMBO-NET: Multi-Causal Aware Modeling Backdoor-Intervention Optimization for Medical Image Segmentation Network}

\author[label1,label2,label3,label4]{Ruiguo Yu} 
\ead{rgyu@tju.edu.cn}
\author[label4]{Yiyang Zhang} 
\ead{zyy\_0203@tju.edu.cn}
\author[label1,label2,label3]{Yuan Tian} 
\ead{tiany@tju.edu.cn}
\author[label1,label2,label3]{Yujie Diao} 
\ead{yujiediao@tju.edu.cn}
\author[label1]{Di Jin}
\ead{jindi@tju.edu.cn}
\author[label6,label7,label8]{Witold Pedrycz\corref{cor1}} 
\ead{wpedrycz@ualberta.ca}
\cortext[cor1]{Corresponding author}
\fntext[equal]{These authors contributed equally to this work.}

\affiliation[label1]{organization={College of Intelligence and Computing, Tianjin University}, 
            city={Tianjin},
            postcode={300350}, 
            country={China}}
\affiliation[label2]{organization={Tianjin Key Laboratory of Cognitive Computing and Application},
            city={Tianjin},
            postcode={300350}, 
            country={China}}
\affiliation[label3]{organization={Tianjin Key Laboratory of Advanced Networking},
            city={Tianjin},
            postcode={300350}, 
            country={China}}
\affiliation[label4]{organization={Tianjin International Engineering Institute, Tianjin University},
            city={Tianjin},
            postcode={300350}, 
            country={China}}

\affiliation[label5]{organization={Tianjin Central Hospital of Gynecology Obstetrics},
            city={Tianjin},
            postcode={300100}, 
            country={China}}
              
\affiliation[label6]{organization={Department of Measurement and Control Systems, Silesian University of Technology, Akademicka 2 Gliwice, 44-100}, country={Poland}}

\affiliation[label7]{organization={Department of Electrical and Computer Engineering, University of
Alberta, Edmonton, AB, T6G 2R3}, country={Canada}}

\affiliation[label8]{organization=
{Research Center of Performance and Productivity Analysis, Istinye University, Istanbul, 34010}, country={Turkiye}}

%

\begin{abstract}
Medical image segmentation methods generally assume that the process from medical image to segmentation is unbiased, and use neural networks to establish conditional probability models to complete the segmentation task. This assumption does not consider confusion factors, which can affect medical images, such as complex anatomical variations and imaging modality limitations. Confusion factors obfuscate the relevance and causality of medical image segmentation, leading to unsatisfactory segmentation results. To address this issue, we propose a multi-causal aware modeling backdoor-intervention optimization (MAMBO-NET) network for medical image segmentation. Drawing insights from causal inference, MAMBO-NET utilizes self-modeling with multi-Gaussian distributions to fit the confusion factors and introduce causal intervention into the segmentation process. Moreover, we design appropriate posterior probability constraints to effectively train the distributions of confusion factors. For the distributions to effectively guide the segmentation and mitigate and eliminate the impact of confusion factors on the segmentation, we introduce classical backdoor intervention techniques and analyze their feasibility in the segmentation task. Experiments on five medical image datasets demonstrate a maximum improvement of 2.28\% in Dice score on three ultrasound datasets, with false discovery rate reduced by 1.49\% and 1.87\% for dermatoscopy and colonoscopy datasets respectively, indicating broad applicability.
\end{abstract}






\begin{keyword}
Medical Image Segmentation\sep Causal Inference\sep Backdoor Model\sep Gaussian Modeling
\end{keyword}

\end{frontmatter}



\section{Introduction}
\label{sec1}
\begin{figure}[!h]
    \centering
    \includegraphics[scale=1,width=1\linewidth]{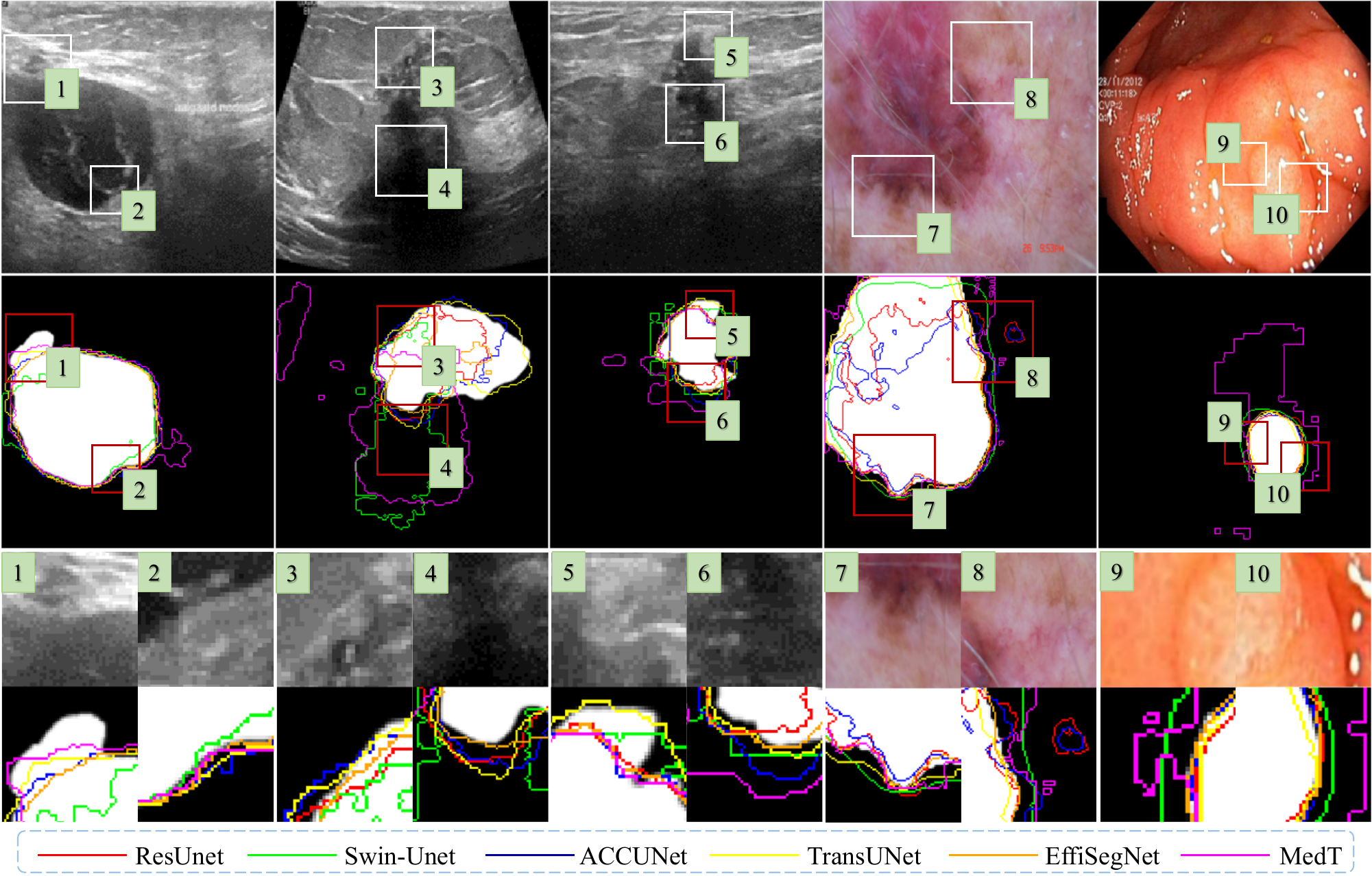}
    \caption{The segmentation results of multiple networks. The third row corresponds to a zoomed-in image of the selected area. The curves of different colors in the second and last rows represent the corresponding models' predictions of the lesion area.}
    \label{fig:clair}
\end{figure}

Medical image segmentation is a vital component of computer-aided diagnosis, playing a key role in assisting clinicians with treatment planning and decision-making. In recent years, deep learning methods such as UNet~\cite{unet} and its variants~\cite{resunet,unetpp,unext}
have successfully been employed in various segmentation tasks in pathology and imaging modalities, including colon polyp segmentation, skin lesion segmentation, and breast nodules analysis. These methods demonstrate the ability to classify regions of interest or lesion pixels accurately. 
However, they still face challenges in achieving high accuracy and reliability in pixel classification at the boundaries of the segmented areas. As illustrated in Fig.~\ref{fig:clair}, different segmentation methods exhibit varying degrees of under-segmentation and over-segmentation at the boundary.

Many existing segmentation methods utilize networks that directly couple feature extraction with segmentation prediction. However, this approach introduces confusion factors where the features become intertwined with target information, resulting in biased model predictions. Causal learning theory~\cite{causal_infer} suggests that modeling and incorporating causal interventions can alleviate the negative impact of confusion factors. For instance, JointMCMC~\cite{JointMCMC} proposes a Monte Carlo sampling embedding model that probabilistically describes noise patterns and reduces speckle noise in ultrasound image lesion segmentation. SwinHR~\cite{SwinHR} demonstrates effective modeling of kinetic features at different acquisition times to attenuate the influence of lesion heterogeneity on segmentation. While causal inference methods help mitigate confusion factors, existing approaches often rely on manual hierarchical modeling~\cite{manual1,manual3} or oversimplify by focusing on specific causal concepts, neglecting other potential critical factors. Additionally, some abstract concepts within the confusion factors (e.g., fuzzy boundaries or underlying pathological features) are complex to describe quantitatively, thus increasing the complexity of modeling.

To comprehensively model confusion factors and mitigate their influence on the segmentation process, we propose the Multi-causal Aware Modeling Backdoor-intervention Optimization for Medical Image Segmentation Network (MAMBO-NET). For complete modeling of confusion factors, MAMBO-NET introduces a latent space modeling approach based on Gaussian self-modeling to address the challenges posed by complex confusion factors. Through intervention optimization, the method effectively reduces the confounding effects on model decision-making, thereby achieving significant improvements in segmentation performance. Specifically, our contribution can be summarised as follows:
\begin{itemize}
    \item We classify the adverse factors that affect segmentation as confusion factors and employ the construction of a structured causal diagram to analyze their underlying mechanisms. 
    \item The implicit modeling approach of abstract and concrete confusion factors using Gaussian Self-modeling is proposed. This approach enables the self-modeling of a more comprehensive latent space of confusion concepts, eliminating the need for dependency on existing manual confusion modeling methods.
    \item To reduce the bias of confusion factors on segmentation decisions, we propose combining implicit modeling with the segmentation process through backdoor interventions.
    \item Our method achieved a maximum segmentation index improvement of 2.28\% on three ultrasound datasets, and reduced lesion area false alarm rates by 1.49\% and 1.87\% in the dermatoscopy and colonoscopy datasets, respectively, indicating that the method has broad applicability.
\end{itemize}

\section{Related Work}
\subsection{Latent Space Modeling}
\label{latentSpaceModeling}
Latent space modeling effectively characterizes high-dimensional lesion features and their implicit relationships to enhance segmentation performance. Current approaches predominantly follow three paradigms: feature decoupling, prototype learning, and probabilistic modeling. Feature decoupling methods, exemplified by KDFD~\cite{KDFD} for content-style separation and CDDSA~\cite{CDDSA} for contrast domain disentanglement, isolate confusion factors from target features, though potential information loss remains a limitation. Li et al.~\cite{li2010l1} proposed a foundational work in image representation, which has been widely used in recent decades. It achieved exceptional robustness in handling outliers, having a profound impact on the development of image processing. 

Prototype-based techniques like SSMIS~\cite{SSMIS} with boundary-aware prototypes and PROCNS~\cite{PROCNS} using progressive calibration offer more compact representations through feature clustering, yet face challenges in capturing complex confusion relationships. Probabilistic approaches, including GMM-SDF~\cite{GMM-SDF} for morphological variations and BayeSeg~\cite{BayeSeg} for domain adaptation, provide flexible distribution modeling, though their potential for segmentation tasks warrants deeper investigation.

\subsection{Medical Image Segmentation Based on Causal Inference}

Recent advances in causal inference have spurred its integration into medical image analysis, primarily focusing on causal interpretation, intervention, and counterfactual prediction.

Causal interpretation methods analyze relationships between anatomical structures and predictions. C-CAM~\cite{C-CAM} constructs interpretable causal models by examining category-anatomical causality, while P-CSS~\cite{P-CSS} models disease co-occurrence effects to mitigate bias in radiology reports.
Causal intervention addresses prediction bias through frontdoor or backdoor adjustments. CaMIL~\cite{Camil} reduces spurious disease-color associations in WSI classification via frontdoor adjustment, whereas CausalCLIPSeg~\cite{CausalCLIPSeg} suppresses confusion bias in segmentation using backdoor adjustment.
Counterfactual prediction frameworks estimate potential outcomes under interventions. Wang et al.~\cite{wang2021bilateral} synthesize counterfactual mammographic features, while Richens et al.~\cite{richens2020improving} demonstrate how counterfactual reasoning improves diagnostic decisions.
Medical images contain entangled confusion factors, complicating their separation in high-dimensional feature space. We propose a latent space modeling approach with backdoor adjustment to mitigate their adverse effects on segmentation.

\section{Method}
\subsection{MAMBO-NET Architecture}
\begin{figure}[t]
    \centering
    \includegraphics[scale =1,width=1\linewidth]{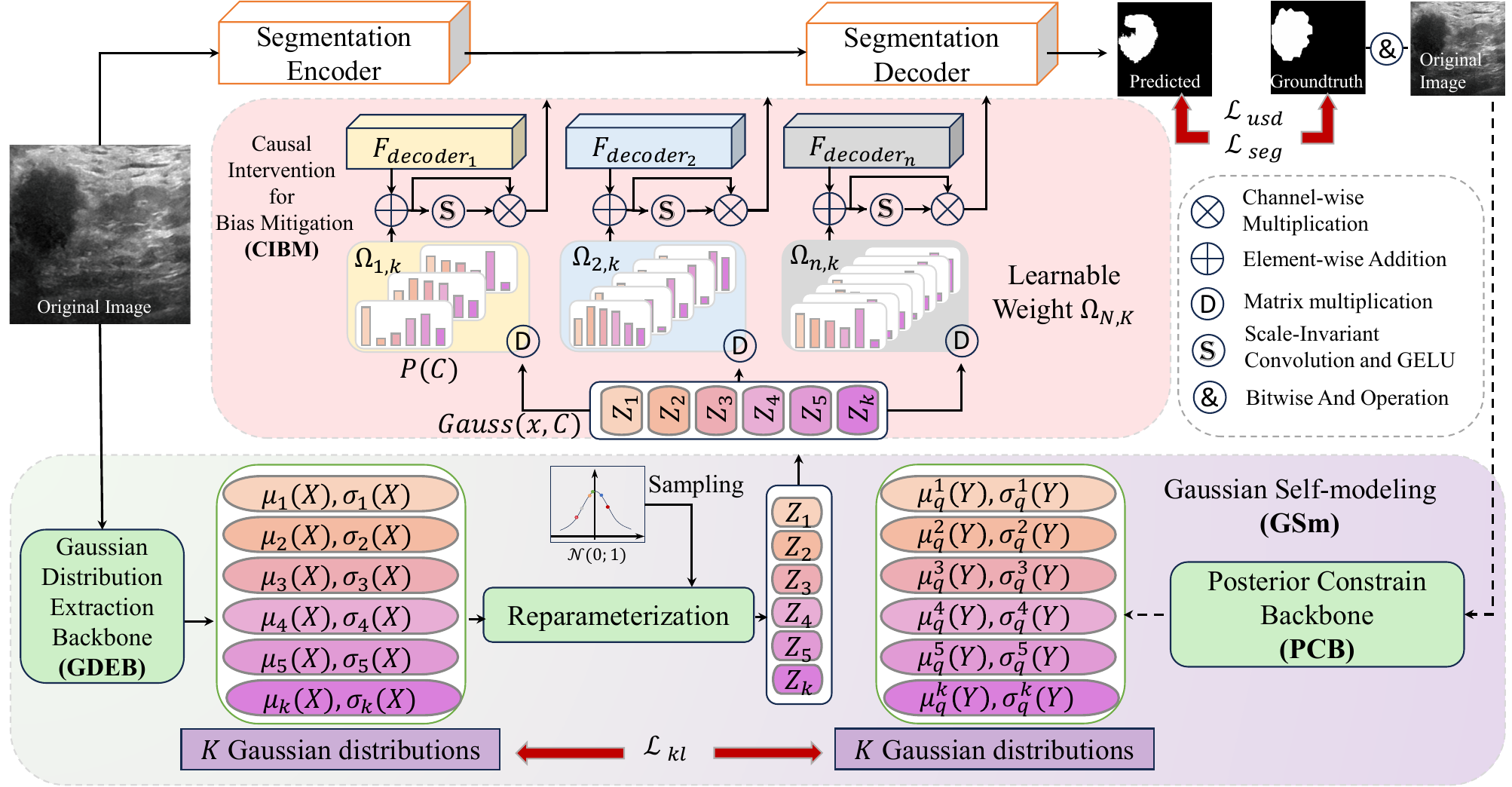}
    \caption{The architecture of the proposed \textbf{MAMBO-NET}. Dashed lines indicate that the data stream is disabled during model inference. \textbf{Segmentation Encoder} and \textbf{Segmentation Decoder} are the encoder and the decoder in UNeXt. \textbf{Gaussian Backbone} and \textbf{Posterior Constrain Backbone} will use the global average pooling(GAP) and linear mapping for scale alignment. }
    \label{arch}
\end{figure}
MAMBO-NET proposes a causal intervention framework with three core components: causal relationship modeling, Gaussian-based implicit modeling (GSm), and bias-aware intervention (CIBM). As illustrated in Fig.~\ref{arch}, the architecture integrates: (1) the UNeXt \cite{unext} segmentation backbone(Encoder and Decoder), (2) GSm module with Gaussian Distribution Extraction Backbone (GDEB) and posterior constraint backbone (PCB) for confusion factors modeling, and (3) CIBM for intervention fusion. GSm employs reparameterization to sample prior distributions while PCB provides training-phase constraints. CIBM performs weighted fusion of bias-mitigated features across decoder stages. The framework implements backdoor adjustment to suppress confusion effects through: explicit causal modeling of hidden confusion factors (Sec.~\ref{crm}), GSm's learnable Gaussian priors with posterior regularization (Sec.~\ref{gsm}), and CIBM's adaptive feature conditioning (Sec.~\ref{cibm}).

\subsection{Causal Relationship Modeling}
\label{crm}
The backdoor model is one of the most classical models in causal relationships. Previous works have attempted to incorporate the backdoor model into medical imaging tasks and model and intervene in the confusion factors that affect model decisions, such as imaging artifacts and scattered noise. The modeling of confusion factors in these works is based on hierarchical assumptions made by humans about known concepts. Therefore, the confusion factors usually only include visible and concrete concepts. However, in this paper, we unify the known and unknown, abstract and concrete factors that affect segmentation boundary decisions into confusion factors, which are distributed at the segmentation boundaries. For example, the comet tail sign in ultrasound images is a known, concrete concept, while factors such as physician acquisition habits are unknown, abstract concepts.

\begin{figure}[ht]
    \centering
    \includegraphics[width=1\linewidth]{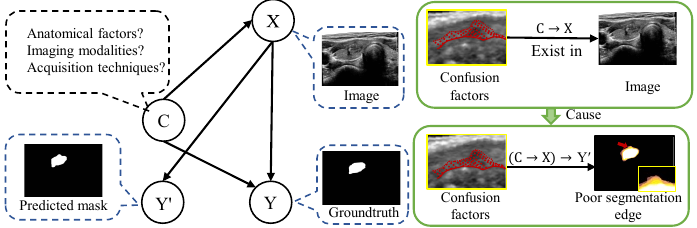}
    \caption{The process of causal relationship modeling. X denotes the original image; Y denotes the predicted mask, and c denotes the confusion factors.}
    \label{fig:causal-manbo}
\end{figure}

Fig.~\ref{fig:causal-manbo} represents the causal backdoor model, where $X$ denotes the input image and $Y^{\prime}$ represents the segmentation mask produced by the network. The variable $C$ represents the confusion factors described in this paper. Since the confusion factors exist within the acquired medical image $X$, we consider that the confusion factors have a causal effect on the acquired medical image $X$. Simultaneously, due to the complex causal relationship between the segmentation mask $Y^{\prime}$ and the complex confusion factors, the confusion factors C also have a causal effect on the segmentation mask $Y^{\prime}$. 
Traditional segmentation models utilize a prediction process where the encoder extracts feature representations $\phi(X; \theta)$ from the medical image $X$, and the decoder generates the corresponding segmentation mask $Y^{\prime}$, i.e., the process of obtaining $P(Y^{\prime}|X)$. However, these models typically overlook the impact of confusion factors $C$. As $C$ becomes the confounder acting on both $X$ and $Y^{\prime}$, the learned $P(Y^{\prime}|X)$ by the network becomes biased.

To mitigate the bias introduced by the confounder C on the network's learned decisions and enforce the network to learn an unbiased decision rule $P(Y^{\prime}|do(X))$, where the do operator represents causal intervention, we aim to adjust the backdoor using a hierarchical modeling of the confusion factors $C$, which can be expressed as Eq.~\eqref{intervention}.
\begin{equation}
    P(Y^{\prime}|do(X)) = \sum_{c_i}^C P(Y^{\prime}=y^{\prime}|X=x,C=c_i)P(C=c_i)
    \label{intervention}
\end{equation}
where $x$ represents the current image characteristics and $y^{\prime}$ represents the prediction result of the network. $c_i$ is a concept of the set of confusion factors in the current image.

\subsection{Gaussian-based implicit modeling}
\label{gsm}
The confusion factors summarized by existing methods often only cover observable and known concepts, oversimplifying the nature of the confusion factors. Since the confusion factors we define encompass a broader range of abstract concepts, we introduce a Gaussian mixture model to represent the Gaussian distributions of the generalized and unknown abstract concepts.

In the Gaussian Self-Modeling (GSm) module, we associate the features corresponding to the potential $K$ types of confusion concepts with $K$ Gaussian distributions in Eq.~\eqref{equ:p-gaussian}.
\begin{equation}
 \mathcal{N}(\mu_P (X;\theta) , {\sigma_P}^2(X;\theta) )= {\left\{ \mathcal{N}(\mu_p
^{i}( X;\theta) , {\sigma_p^{i}}^2( X;\theta)) \right\} }_{i=1}^K
    \label{equ:p-gaussian}
\end{equation}
where $\mathcal{N}(\mu_p
^{i}( X;\theta) , {\sigma_p^{i}}^2( X;\theta))$represents the prior Gaussian distribution of the $i$-th group of confusion factors, $\mu_p
^{i}$ and $\sigma_p^{i}$ represent the mean and standard deviation of the Gaussian distribution.

Since the number of confusion factors varies across tasks, we avoided manually setting concept layers and instead used a large constant KK, allowing the network to learn relevant concepts implicitly. Unlike existing methods that average concept features, treating confounders as static, our approach accounts for their randomness and variability across samples.

To maintain the stochasticity and variability, we utilize the Gaussian distribution sampling strategy to represent statistical information while modeling the statistical characteristics of the latent space for each confusion factors. Moreover, we utilize the reparameterization technique\cite{vae} to make distribution sampling models trainable. Specifically, the sampled feature represented by the $i$-th Gaussian distribution can be expressed as Eq.~\eqref{sample}.
\begin{equation} 
z_i =  \epsilon \cdot \sigma_{i}( \textit{X};\theta) + \mu_{i}( \textit{X};\theta).
\label{sample}
\end{equation} 
where $ \epsilon $ is sampled from the standard normal distribution $\epsilon \sim \mathcal{N}(0, 1) $ and $\mu_{i}, \sigma_{i}$ are the mean and standard deviation extracted from the GDEB.

To utilize posterior information to constrain the Gaussian modeling process, we construct a posterior constraint backbone by using the real mask $Y$ corresponding to the input image $X$ to learn the distribution space of confusion factors. We employ another backbone network to extract the posterior probability distribution regarding $K$ confusion factors, which can be expressed as Eq.~\eqref{equ:Q-gaussian}.
\begin{equation}
     \mathcal{N}(\mu_{Q}( \textit{Y};\theta^{\prime}), {\sigma_{Q}}^2( \textit{Y};\theta^{\prime}) )= {\left\{ \mathcal{N}(\mu_{q}^{i}( \textit{Y};\theta^{\prime}), {\sigma_{q}^{i}}^2( \textit{Y};\theta^{\prime})) \right\} }_{i=1}^K,
    \label{equ:Q-gaussian}
\end{equation}
where $\mathcal{N}(\mu_q^{i}( X;\theta) , {\sigma_q^{i}}^2( X;\theta))$represents the posterior Gaussian distribution of the $i$-th group of confusion factors, $\mu_q
^{i}$ and $\sigma_q^{i}$ represent the mean and standard deviation.

We utilize $KL$ divergence to minimize the distance between the prior probability distribution of confusion factors and the posterior-constrained probability distribution. $KL$ divergence loss is designed for multiple independent Gaussian distributions, which can be expressed as Eq.~\eqref{losskl}.
\begin{equation}
\begin{aligned}
    \mathcal{L}_{kl} &= KL(\mathcal{N}(\mu_P,\sigma_P^2) \parallel \mathcal{N}(\mu_Q,\sigma_Q^2) ) \\
    &= \frac{1}{K} \sum_{i=1}^K  \left [  \log{\dfrac{\sigma_q^i}{\sigma^i}}+\dfrac{{\sigma^i}^{2}+(\mu^i-\mu_q^i)^2}{2{\sigma_q^i}^{2}}-\dfrac{1}{2}  \right ],
\label{losskl}
\end{aligned}
\end{equation}
where $ \mu^i , \sigma^i $ and $ \mu_q^i , \sigma_q^i$ represent the $i$-th set mean and standard deviation in $K$-set distribution features. 

To add strong constraints to the distribution from the segmentation process and to limit the modeling space of the confusion factors, we only focus on the segmentation boundary regions that are susceptible to bias interference. To achieve this, we propose $\mathcal{L}_{usd}$ that limits the uncertainty spatial distribution, which can be expressed as Eq.~\eqref{lossusd}.
\begin{equation}
    \mathcal{L}_{usd} = - \frac{1}{N} \displaystyle\sum\nolimits_{i=1}^{N}(1+V_i)\cdot[b_{i}log\hat{b}_{i}  + (1-b_{i})log(1-\hat{b}_{i})],
    \label{lossusd}
\end{equation}
where $b_i$ and $\hat{b}_{i}$ denote the ground truth and predicted values of the pixel $i$ within within the range of the boundary. To compute $b_i$, we extract the edge regions of the mask using the Sobel operator in our implementation. $V_i$ represents the degree of uncertainty of pixel $i$. We calculate the mean value $P$ of the pixels and compute the variance $V_i$ to quantify the uncertainty of the pixel $i$. This process can be described as Eq.~\eqref{uusd}.
\begin{equation}
    V_i=(p_i-P)^2, \enspace where \enspace P=\frac{1}{N} \displaystyle\sum\nolimits_{i=1}^{N} p_{i},
    \label{uusd}
\end{equation}
where $N$ is the number of pixels, $p_i$ is the predicted pixel value before the thresholding process. Finally, we define the loss for implicit modeling of confusion factors as  Eq.~\eqref{lossgaus}.
\begin{equation}
    \mathcal{L}_{Gaus}=\mathcal{L}_{usd}+\mathcal{L}_{kl},
\label{lossgaus}
\end{equation}

\subsection{Bias-Aware Causal Intervention}
\label{cibm}
To implement an active intervention strategy for confusing features during segmentation, the CIBM is proposed, which integrates confusion semantics and decoding semantics to achieve bias intervention of confusion factors on prediction results. CIBM aims to implement the confusion factors intervention term $P(Y^{\prime} | do(X))$ in Eq.~\eqref {intervention}. CIBM uses a normalized weighted geometric mean to project the probability prediction of confusion factors in the image onto the semantic feature space, which can be expressed as Eq.~\eqref{equ:tuidao3}.
\begin{equation}
    \begin{aligned}
        P(Y^{\prime}\mid do(X)) & = \mathbb{E}_{\hat{C}\sim P(\hat{C} \mid X)}[P(Y^{\prime} \mid X,\hat{C})] \\ 
    & \approx P(y^{\prime}\mid x,\sum\nolimits_{\hat{c}_i}^{\hat{C} }\hat{c}_iP(\hat{c}_i))\\
    & \approx P(y^{\prime}\mid x \odot \sum\nolimits_{\hat{c}_i}^{\hat{C}} (Gauss(x, \hat{c}_i;\theta^{\prime\prime}) \cdot P(\hat{c}_i))),
    \end{aligned}
    \label{equ:tuidao3}
\end{equation}
where $\odot$ denotes the fusion operation of confusion factors modeling and original image mapping. $Gaus(x, c_i; \theta^{\prime\prime})$ represents the implicit modeling of the confusion factors $c_i$, which appearing in the image $x$.

CIBM assumes that the occurrence of various confusion factors follows a learnable random variable distribution. This modeling assumption is based on class differences observed in actual scenarios, such as in thyroid nodule imaging, where there is no equal relationship between edge blur and the frequency of specular reflection artifacts. CIBM introduces a learnable probability parameter $\Omega$, which is used to construct Gaussian mixture distributions and control the contribution strength of each confusion factor to semantic features as a mixing coefficient. The weighted modeling of the Gaussian mixture distribution is shown in Eq.~\eqref{equ:mixgauss3}.
\begin{equation}
     \sum\nolimits_{\hat{c}_i}^{\hat{C}} (Gauss(x, \hat{c}_i;\theta^{\prime\prime}) \cdot P(\hat{c}_i)) = \Omega \times Z,\enspace s.t. \displaystyle\sum\nolimits_{k=1}^{K}\Omega_{n, k}=1
     \label{equ:mixgauss3}
\end{equation}
where $\Omega \in \mathbb{R}^{n \times K}$ represents the weighted probability value of $K $ dimensional confusion features, $Z \in \mathbb{R}^{K}$ is the set of confusion features sampled in Eq.~\eqref{sample}, and $n$ represents the number of feature channels corresponding to the decoding stage. The matrix multiplication operation $\Omega \times Z$ achieves feature fusion of confusion semantics, allowing the features of each channel in the corresponding decoding stage to adaptively adjust the interference level of different confusion factors.

The semantic feature $\Omega \times Z$ in the CIBM module is replicated and expanded into a two-dimensional feature map through the channel dimension, and is matched with the corresponding decoding feature $F_{ {decoder}_i }$ to perform cascading splicing. Splicing features utilize convolution and Gaussian Error Linear Unit (GELU) based on Gaussian distribution to enhance expressive power, in the form of channel weights $\mathbb{S}$, compared with the original decoding feature $F_{{decoder}_i}$ performs channel wise weighting to form an enhanced decoding output, and the fusion calculation method is as Eq.~\eqref{equ:fusion3}.
\begin{equation} 
F = \mathbb{S} \otimes  \left ( F_{{decoder}_i} \oplus \text{Repeat}(\Omega \times Z) \right )
\label{equ:fusion3} 
\end{equation}
where $\oplus$ represents element wise addition, $\otimes$  is channel level multiplication operation, $\mathbb{S}$ represents the channel weight of the fused feature after convolution and activation operation, and $\text{Repeat}(\cdot)$ represents expanding the one-dimensional vector in the spatial dimension into a two-dimensional feature matrix that matches $F_{{decoder}_i}$.

The MAMBO-NET model guides end-to-end network learning through multiple loss functions. For image segmentation tasks, a combination of Binary Cross Entropy Loss (BCELoss) and Dice loss is used to simultaneously optimize pixel level classification accuracy and region overlap; For the modeling task of confusion factors bias intervention, the Eq.\eqref{losskl} and \eqref{lossusd} are used to model the confusion factors loss, guiding the model to learn the distribution characteristics of the confusion factors and optimize the segmentation ability of the confusion factors region.
BCELoss improves pixel-level classification accuracy by calculating the difference between the predicted mask and the true mask pixel by pixel. The calculation of BCELoss is shown in Eq.~\eqref{equ:bceloss3}.
\begin{equation}
     L_{BCE} = -\frac{1}{N} \displaystyle\sum\nolimits_{i=1}^{N}[y_{i}log\hat{y}_{i}  + (1-y_{i})log(1-\hat{y}_{i})]
     \label{equ:bceloss3}
\end{equation}
where $y_i $ represents the true mask value of the $i$ th pixel, $\hat {y}_i $ represents the predicted value of the $i $th pixel, and $N $ is the number of pixels.

Dice loss measures segmentation performance from the perspective of region overlap and spatial consistency between predicted and annotated regions, which is shown as Eq.~\eqref{equ:diceloss3}, and the meaning of the variables is the same as Eq.~\eqref{equ:bceloss3}.
\begin{equation}
     L_{Dice} = 1 - \frac{2\displaystyle\sum\nolimits_{i=1}^{N}y_{i}\hat{y}_{i}}{\displaystyle\sum\nolimits_{i=1}^{N}y_{i} + \displaystyle\sum\nolimits_{i=1}^{N}\hat{y}_{i}}
     \label{equ:diceloss3} 
\end{equation}
The total loss of the MAMBO-NET model consists of segmentation loss and confusion factors modeling loss, and the total loss expression is shown in Eq.~\eqref{equ:totalloss3}, where $L_{Gaus}$ is Eq.~\eqref{lossgaus}. 
\begin{equation}
     L_{Total} =  L_{Gaus} +  L_{BCE} + L_{Dice}
     \label{equ:totalloss3}
\end{equation}

\section{Experiments}
\subsection{Datasets and Settings} 
\textbf{Datasets.} We evaluated on five medical datasets: BUSI~\cite{BUSI} (665 breast ultrasound images), DDTI~\cite{DDTI} (3,644 thyroid ultrasound images), TUI~\cite{TUI} (15,233 thyroid ultrasound images), ISIC2018~\cite{ISIC2018} (3,694 dermoscopy images), and KVASIR~\cite{KVASIR} (1,000 colonoscopy images).

\textbf{Settings.} Using PyTorch on RTX 3090, we trained with SGD (momentum=0.9, weight decay=0.01), lr=1e-3, batch=32 for 200 epochs with cosine decay. Data was split 70\%/30\% (train/test) with standard augmentation (flipping/rotation/cropping).

\subsection{Comparison with state-of-the-art methods}


\begin{table}[!t]
\centering
\caption{The experimental results were obtained from three ultrasound datasets. The upward arrow $\uparrow$ denotes that a higher value is preferable, whereas the downward arrow $\downarrow$ indicates that a lower value is preferable. The best performance is highlighted in bold, and a horizontal line is used to mark the second-best performance.} 
\label{res-eval-three} 
\resizebox{\linewidth}{!}{
\begin{tabular}{ccccc||cccc||cccc}
\toprule 
 \multirow{2}{*}{Methods}  & \multicolumn{4}{c}{BUSI}          & \multicolumn{4}{c}{DDTI}  & \multicolumn{4}{c}{TUI}         \\ 
  \cmidrule(r){2-5} \cmidrule(r){6-9}\cmidrule(r){10-13}
  &  Dice $\uparrow$    
  &  IoU $\uparrow$  
  &  FDR $\downarrow$   
  &  AUC $\uparrow$    
   &  Dice $\uparrow$    
  &  IoU $\uparrow$  
  &  FDR $\downarrow$   
  &  AUC $\uparrow$ 
   &  Dice $\uparrow$    
  &  IoU $\uparrow$  
  &  FDR $\downarrow$   
  &  AUC $\uparrow$ \\ \midrule

UNet\cite{unet}(2015)    & 66.67 & 50.01 & 32.41 & 91.41 & 79.99 & 66.92 & 25.47 & 96.49  & 87.00 & 77.20 & 12.03 & 97.83\\
UNet++\cite{unetpp}(2018)   & 74.44 & 59.35 & 20.00 & 91.07 & 84.29 & 72.99 & 13.36 & 96.89 & 89.01  & 80.34 & 8.56 & 97.73\\
ResUNet\cite{resunet}(2020)    & 70.69 & 55.17 & 44.86 & 92.26 & 82.94 & 71.21 & 53.61 & 95.11 & 86.31 & 76.31  & 17.15   & 95.18\\
MedT\cite{medt}(2021)  &  67.90  &  51.55 &  31.83 &  94.21 &  73.66 &  58.73 &  25.93 &  97.61 &  77.52 &  63.82 &  20.94 &  98.38\\
GGNet\cite{ggnet}(2021) &  70.20  &  54.10  &  29.92 &  93.98 &  71.54 &  55.94 &  28.45 &  96.88 &  80.39 &  67.44 &  17.66 &  97.71\\
TransUNet\cite{transunet}(2021) &  74.48 &  59.66 &  21.45 &  93.76 &  86.25 &  76.12 &   12.48 &  98.33 &  89.09 &  80.38 &  10.91 &  98.90\\
AAU-net\cite{aaunet} (2022)   &  \underline{77.14} &  \underline{62.86} &  19.48 & 91.47 & 87.04 & 77.31 &  12.74 &  98.41 & 87.54  & 78.02  & 12.14  & 98.83\\
Swin-Unet\cite{swinunet}(2022) &  65.03 &  48.50  &  33.28 &  93.80  &  66.02 &  49.82 &  35.54 &  96.04 &  56.36 &  39.78 &  44.79 &  95.44\\
UNeXt \cite{unext} (2022)                     & 74.28 & 59.19 & 22.98 & 93.31 &  87.22 &  77.46 & 24.92 & 95.24 &  91.04 &  83.58 &  \underline{8.50} & 99.46\\
Acc-Unet\cite{accunet} (2023)  & 74.63 & 59.89 &  \textbf{16.53} &  93.79 & 85.41 & 74.83 &  \textbf{12.21} &  \underline{98.69} & \underline{89.94}  & \underline{82.55}  & 10.68  &  99.56\\
BUSSeg\cite{busseg} (2023)& 75.24&60.51&18.49&93.61 &\underline{87.38}&\underline{77.63}&13.46&97.62&89.23 &80.78 &9.27&\textbf{99.76}\\
EffiSegNet\cite{EffiSegNet}(2024) & 73.45 & 58.22 & 20.80  & \underline{94.35} & 84.56 & 73.42 &  13.87 &  98.34 & 88.39  & 79.43  & 11.44  & 99.29  \\
LMNet\cite{LM-Net} (2024)     & 74.31 & 59.28  & 26.02 &  94.01 & 83.05 & 71.27 & 16.38 & 94.29   & 86.38  & 76.71  & 21.72  & 98.37\\
MAMBO-NET(OURS)  &   \textbf{77.94}   &   \textbf{64.40}   &   \underline{18.36}   &   \textbf{94.70}   &   \textbf{89.66}   &  \textbf{81.90}   &  \underline{12.50}   & \textbf{98.72}  & \textbf{91.10}      &   \textbf{83.96}      &  \textbf{7.95 }     &  \underline{99.63}   \\ \bottomrule 
\end{tabular}
}
\end{table}

\begin{table}[!t]
\centering
\caption{The experimental results on ISIC2018 and KVASIR datasets. The best performance is highlighted in bold, and a horizontal line is used to mark the second-best performance.} 
\label{res-eval-two} 
    \resizebox{\linewidth}{!}{
    \begin{tabular}{ccccc||cccc}
    \toprule 
    & \multicolumn{4}{c}{ISIC2018} & \multicolumn{4}{c}{KVASIR} \\ \cmidrule(r){2-5} \cmidrule(r){6-9}
    & Dice $\uparrow$   
    & IoU $\uparrow$  
    & FDR $\downarrow$
    & AUC $\uparrow$  
    & Dice $\uparrow$   
    & IoU $\uparrow$
    & FDR $\downarrow$
    & AUC$\uparrow$  \\ \midrule
    UNet\cite{unet}(2015)  & 86.21 & 76.12 &15.22&95.83& 75.63& 61.27 &30.25&95.02     \\
    UNet++ \cite{unetpp}(2018)& 88.65 & 79.52&21.01&95.37& 81.87& 69.35&18.99&95.21      \\
    ResUNet \cite{resunet}(2020)      & 88.14 & 79.03&19.83&96.78& 73.98& 58.97 &22.71&92.76     \\
    MedT  \cite{medt}(2021)& 87.69 & 78.37&17.22&96.31& 62.18& 45.44 &37.44&91.79     \\
    GGNet \cite{ggnet}(2021)& 87.91 & 78.54&14.24&95.27& 68.83& 52.55  &33.54&95.81    \\
    TransUNet \cite{transunet}(2021)& 88.71 & 79.84&25.64&94.28& 83.22& 71.36  &16.75&93.92    \\
    AAU-Unet  \cite{aaunet}(2022)    & 89.16 & 80.47&12.98&96.36& 82.87& 71.06  &19.23&96.39    \\
    Swin-Unet \cite{swinunet}(2022)    & 87.53 & 78.14&16.47&94.43& 61.98& 45.13  &40.22&92.83    \\
    UNeXt \cite{unext}(2022)& 88.66 & 79.65&19.73&\textbf{97.71}& 81.71& 68.22  &\underline{12.11}&96.33    \\
    Acc-unet \cite{accunet}(2023)      & 85.67 & 75.31&14.39&92.74& 79.49& 65.72    &23.21&95.94  \\
    BUSSeg \cite{busseg} (2023)&86.53 & 76.84 & 15.37&95.27 & 80.19 &67.01  &18.24 &\underline{96.77} \\
    EffiSegNet \cite{EffiSegNet}(2024) &  \underline{89.26} & \underline{80.85} & \underline{12.88}&95.94 & \textbf{84.27}  & \textbf{72.98}&13.49&93.28  \\
    LMNet\cite{LM-Net}(2024)      & 87.50  & 77.87 &17.47&95.32 & 80.38  & 67.24 &24.28&91.54 \\
    MAMBO-NET(OURS) & \textbf{89.30} & \textbf{80.97} &\textbf{11.39}&  \underline{97.23}& \underline{83.96}& \underline{72.11}  &\textbf{10.24}&\textbf{97.88}  \\ \bottomrule  
    \end{tabular}
    }

\end{table}
\begin{figure}[!t]
    \centering
    \includegraphics[width=1\linewidth]{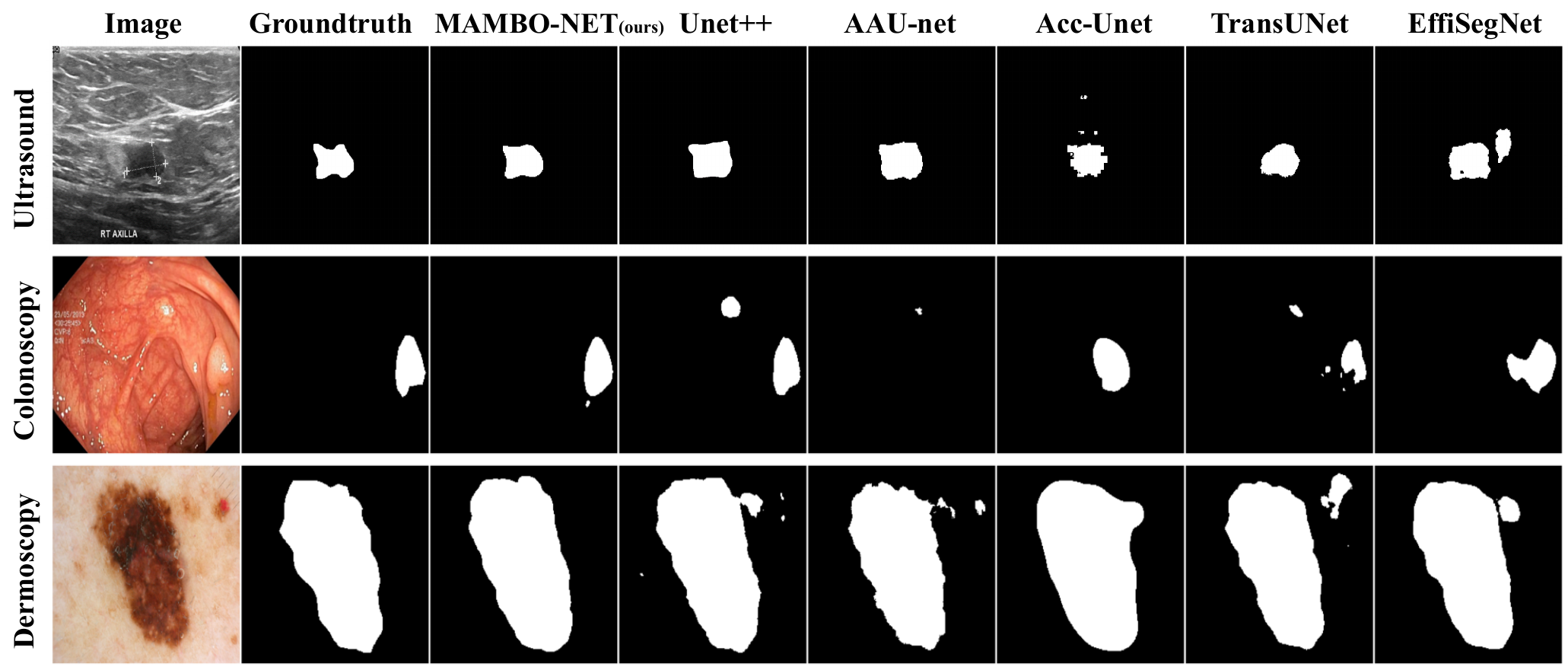}
    \caption{Visualization results of segmentation of multiple models on ultrasound, dermatoscopy, and colonoscopy images.}
    \label{fig:res-little}
\end{figure}

Tab.~\ref{res-eval-three} shows MAMBO-NET's superior performance on ultrasound datasets, outperforming UNeXt by 3.66\% in Dice and 2.33\% in IoU. On BUSI, it improved Dice by 0.8\%, IoU by 1.54\%, and AUC by 0.35\%, while on DDTI, gains were 2.28\% in Dice, 4.27\% in IoU, and 0.03\% in AUC, confirming enhanced segmentation consistency.

Tab.~\ref{res-eval-two} highlights MAMBO-NET's advantages on dermatoscopy and colonoscopy datasets. Compared to UNeXt, it achieved higher Dice (0.64\%) and IoU (1.32\%) with an 8.34\% reduction in FDR. On colonoscopy data, improvements were 2.25\% in Dice, 3.89\% in IoU, and 1.89\% in FDR. Although KVASIR's Dice and IoU were slightly lower than EffiSegNet, it reduced FDR by 3.25\% and improved AUC by 1.11\%, demonstrating better false-positive suppression. Joint analysis reveals MAMBO-NET's particularly significant improvements on low-resolution ultrasound data. Fig.~\ref{fig:res-little} visually confirms segmentation quality enhancement across modalities.


\subsection{Analysis of K sets Gaussian Distributions}

\begin{table}[t]
    \centering
    \scriptsize
    \caption{Ablation experiments on $K$ conducted on the BUSI dataset. The best performance is highlighted in bold. The model uses $K=128$. }
    \begin{tabular}{@{\hspace{6mm}}l@{\hspace{6mm}}c@{\hspace{6mm}}c@{\hspace{6mm}}c@{\hspace{6mm}}c@{\hspace{6mm}}}
    \toprule 
     $K$ & Dice $\uparrow$ & IoU  $\uparrow$ & FDR  $\downarrow$ & AUC $\uparrow$ \\ 
    \midrule
    16       &  76.72&  62.51  &  24.15  &  94.04  \\ 
    32       &  77.43&  63.74  &   \underline{20.46}  &  94.22 \\ 
    128      &   \underline{77.94}&   \underline{64.49}  &  \textbf{ 18.36}  &   \textbf{94.70}      \\ 
    512      &   \textbf{77.98}&  \textbf{64.60 } &  20.74  &  \underline{94.67}      \\ 
    \bottomrule 
    \end{tabular}
    \label{tab:abl-k}
\end{table}
\begin{figure}[t]
    \centering
    \includegraphics[scale = 0.7]{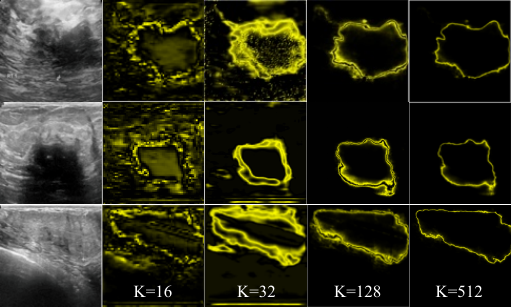}
    \caption{Feature entropy map generated by the decoder layer, where $K$ represents the number of distributions in the GSm.}
    \label{img:abl-un}
\end{figure}


As shown in Tab.~\ref{tab:abl-k}, the results show a non-linear relation between the Gaussian distribution quantity K and model performance. Increasing K from 16 to 128 boosts segmentation metrics, but expanding to 512 causes saturation: Dice declines by 0.04\% and IoU by 0.11\% before leveling off, while FDR rises 2.38\%, indicating diminishing returns from excessive components and feature interference. Decoding feature entropy visualizations confirms K's role in modulating feature bias.

Fig.~\ref {img:abl-un} shows that lighter areas mean higher pixel uncertainty. At $K=16$, large light regions in the entropy map suggest poor encoding of confounders, leading to high decoding uncertainty, especially in artifact areas. As K rises to 32 and 128, light areas shrink, showing better confounder representation and reduced uncertainty. At $K=512$, uncertain regions over-converge, implying noise in bias modeling and loss of confounder simulation ability.

\subsection{Ablation of GSm and CIBM}
Two control schemes were set up in the experiment: (1) feature concatenation and linear mapping were used to replace the Causal Intervention for Bias Mitigation(CIBM); (2) replace the sampling features of GSm with decoder features. The module ablation experiment will be conducted on the BUSI dataset, and the experimental results are shown in Tab.~\ref{tab:ablbusimodule}. Results demonstrate that GSm alone improved Dice by 2.63\ while reducing FDR by 4.22\%, though it caused a slight 1.05\% AUC drop due to increased feature complexity. CIBM alone enhanced Dice by 2.08\% and IoU by 2.43\% while lowering FDR by 3.51\%, but its effectiveness was constrained by limited confounder modeling. The combined use of both modules achieved optimal performance across all metrics, showing their complementary nature - GSm provides confounder distributions that enable more effective bias intervention through CIBM.

\begin{table}[t]
\centering
\scriptsize
\caption{The ablation experiment results of GSm and CIBM(\%)}
\label{tab:ablbusimodule}
    \begin{tabular}{ c c c c c }
        \toprule 
            & Dice $\uparrow$ &IoU  $\uparrow$& FDR  $\downarrow$  & AUC  $\uparrow$  \\ \midrule
        Backbone  & 74.28 & 59.19  & 22.98  &  \underline{93.31}  \\
        Backbone $+$ GSm &  \underline{76.91} &  \underline{62.79}  &  \underline{18.76}  & 92.26  \\
        Backbone $+$ CIBM  & 76.20 & 61.62  & 19.47  & 92.69 \\
        Backbone $+$ GSm $+$ CIBM &  \textbf{77.94}  &  \textbf{64.49} &  \textbf{18.36} & \textbf{94.70}\\  
        \bottomrule 
    \end{tabular}
\end{table}

\subsection{Model selection for GDEB and PCB in GSm}
The VGG16BN\cite{vgg16}, DenseNet\cite{densenet}, and ResNet\cite{resnet} were used as alternative solutions for Gaussian Distribution Extraction Backbone(GDEB) and Posterior Constraint Backbone(PCB). The pre-training involved in the experiment was conducted on ImageNet21k, and global average pooling and linear mapping of the same dimension were used to ensure size alignment across features. The impact of the selection of GDEB and PCB on the model results on the BUSI dataset is shown in Tab.\ref{tab:abl-module}. 
\begin{table}[t]
\centering
\caption{The results of the model selection for GDEB and PCB (\%)}
\label{tab:abl-module}
\resizebox{\linewidth}{!}{
\begin{tabular}{cccccc}
\toprule 
GDEB & PCB  & Dice $\uparrow$ &IoU $\uparrow$& FDR $\downarrow$  & AUC $\uparrow$\\ 
\midrule
VGG16BN(w/o pre) & VGG16BN(w/o pre)   &  \underline{76.63} &  \underline{62.21}  &  20.11  &  \underline{93.87} \\ 
VGG16BN(pre) & VGG16BN(w/o pre)&  76.36   &  61.77  &  \textbf{17.29}  &  92.07   \\ 
VGG16BN(pre) & VGG16BN(pre)&  \textbf{77.94} &  \textbf{64.49}     &  18.36  &  \textbf{94.70}      \\ 
DenseNet(pre) & DenseNet(pre)&  76.22   &  61.64  &  \underline{17.38}  &  93.33 \\
ResNet(pre) & ResNet(pre) &  76.01   &  61.44  &  18.56  &  91.19 \\ 
\bottomrule 
\end{tabular}
}
\end{table}

Both GDEB and PCB used the untrained VGG16BN to achieve a Dice of 76.63\% and an IoU of 62.21\%. However, when only PCB used pre-trained VGG16BN, Dice and IoU decreased by 0.27\% and 0.44\%, respectively. However, when both GDEB and PCB used pre-trained VGG16BN, Dice, IoU, and AUC reached their optimal levels, with improvements of 1.31\%, 2.28\%, and 0.83\% compared to suboptimal levels. Experiments on VGG16BN have shown that ensuring that the prior and posterior constraint distributions are in the same feature space is beneficial for the convergence of KL divergence loss and alignment of distribution. A symmetric architecture design is advantageous for optimizing end-to-end segmentation. The performance of the pre-trained backbone of DenseNet and ResNet lags behind that of VGG16BN, indicating that the architecture of VGG16BN is more suitable for Gaussian feature extraction and posterior constraint modeling.

\section{Conclusion}
We propose MAMBO-NET, a novel framework for medical image segmentation that systematically addresses the challenges posed by confusion factors using causal inference theory. By integrating causal relationship modeling, Gaussian Self-modeling, and causal intervention for bias mitigation, MAMBO-NET effectively disentangles confusion factors from target features and mitigates their impact on segmentation decisions. Experimental results demonstrate its superiority over state-of-the-art methods, with Gaussian Self-modeling proving crucial for comprehensive latent space representation. The further analysis highlights the importance of posterior constraints and pretrained backbones in optimizing performance. These contributions establish MAMBO-NET as an effective approach for enhancing segmentation accuracy and robustness in complex medical imaging scenarios.

\bibliographystyle{elsarticle-num} 

\begin{thebibliography}{10}
\expandafter\ifx\csname url\endcsname\relax
  \def\url#1{\texttt{#1}}\fi
\expandafter\ifx\csname urlprefix\endcsname\relax\def\urlprefix{URL }\fi
\expandafter\ifx\csname href\endcsname\relax
  \def\href#1#2{#2} \def\path#1{#1}\fi

\bibitem{unet}
O.~Ronneberger, P.~Fischer, T.~Brox, U-net: Convolutional networks for biomedical image segmentation, in: Medical image computing and computer-assisted intervention--MICCAI 2015: 18th international conference, Munich, Germany, October 5-9, 2015, proceedings, part III 18, Springer, 2015, pp. 234--241.

\bibitem{resunet}
Z.~Zhang, Q.~Liu, Y.~Wang, Road extraction by deep residual u-net, IEEE Geoscience and Remote Sensing Letters 15~(5) (2018) 749--753.

\bibitem{unetpp}
Z.~Zhou, M.~M. Rahman~Siddiquee, N.~Tajbakhsh, J.~Liang, Unet++: A nested u-net architecture for medical image segmentation, in: Deep Learning in Medical Image Analysis and Multimodal Learning for Clinical Decision Support: 4th International Workshop, DLMIA 2018, and 8th International Workshop, ML-CDS 2018, Held in Conjunction with MICCAI 2018, Granada, Spain, September 20, 2018, Proceedings 4, Springer, 2018, pp. 3--11.

\bibitem{unext}
J.~M.~J. Valanarasu, V.~M. Patel, Unext: Mlp-based rapid medical image segmentation network, in: International conference on MICCAI, Springer, 2022, pp. 23--33.

\bibitem{causal_infer}
M.~Hernan, J.~Robins, Causal inference: What if chapman hall/crc, boca raton (2020).

\bibitem{JointMCMC}
N.~Zhao, A.~Basarab, D.~Kouam{\'e}, J.-Y. Tourneret, Joint segmentation and deconvolution of ultrasound images using a hierarchical bayesian model based on generalized gaussian priors, IEEE transactions on Image Processing 25~(8) (2016) 3736--3750.

\bibitem{SwinHR}
Z.~Zhao, S.~Du, Z.~Xu, Z.~Yin, X.~Huang, X.~Huang, C.~Wong, Y.~Liang, J.~Shen, J.~Wu, J.~Qu, L.~Zhang, Y.~Cui, Y.~Wang, L.~Wee, A.~Dekker, C.~Han, Z.~Liu, Z.~Shi, C.~Liang, Swinhr: Hemodynamic-powered hierarchical vision transformer for breast tumor segmentation, Computers in Biology and Medicine 169 (2024) 107939.

\bibitem{manual1}
B.~Liu, D.~Wang, X.~Yang, Y.~Zhou, R.~Yao, Z.~Shao, J.~Zhao, Show, deconfound and tell: Image captioning with causal inference, in: Proceedings of the IEEE/CVF Conference on CVPR, 2022, pp. 18041--18050.

\bibitem{manual3}
X.~Yang, H.~Zhang, J.~Cai, Deconfounded image captioning: A causal retrospect, IEEE Transactions on Pattern Analysis and Machine Intelligence 45~(11) (2021) 12996--13010.

\bibitem{KDFD}
J.~Wang, C.~Zhong, C.~Feng, Y.~Zhang, J.~Sun, Y.~Yokota, Disentangled representation for cross-domain medical image segmentation, IEEE Transactions on Instrumentation and Measurement 72 (2022) 1--15.

\bibitem{CDDSA}
R.~Gu, G.~Wang, J.~Lu, J.~Zhang, W.~Lei, Y.~Chen, W.~Liao, S.~Zhang, K.~Li, D.~N. Metaxas, et~al., Cddsa: Contrastive domain disentanglement and style augmentation for generalizable medical image segmentation, Medical Image Analysis 89 (2023) 102904.

\bibitem{li2010l1}
X.~Li, Y.~Pang, Y.~Yuan, L1-norm-based 2dpca, IEEE Transactions on Systems, Man, and Cybernetics, Part B (Cybernetics) 40~(4) (2010) 1170--1175.

\bibitem{SSMIS}
Y.~Wang, B.~Xiao, X.~Bi, W.~Li, X.~Gao, Boundary-aware prototype in semi-supervised medical image segmentation, IEEE Transactions on Image Processing (2024).

\bibitem{PROCNS}
Y.~Liu, L.~Lin, K.~K. Wong, X.~Tang, Procns: Progressive prototype calibration and noise suppression for weakly-supervised medical image segmentation, IEEE Journal of Biomedical and Health Informatics (2024).

\bibitem{GMM-SDF}
L.~Zhou, L.~Wang, W.~Li, B.~Lei, J.~Mi, W.~Yang, Multi-stage liver segmentation in ct scans using gaussian pseudo variance level set, IEEE Access 9 (2021) 101414--101423.

\bibitem{BayeSeg}
S.~Gao, H.~Zhou, Y.~Gao, X.~Zhuang, Bayeseg: Bayesian modeling for medical image segmentation with interpretable generalizability, Medical Image Analysis 89 (2023) 102889.

\bibitem{C-CAM}
Z.~Chen, Z.~Tian, J.~Zhu, C.~Li, S.~Du, C-cam: Causal cam for weakly supervised semantic segmentation on medical image, in: Proceedings of the IEEE/CVF Conference on CVPR, 2022, pp. 11676--11685.

\bibitem{P-CSS}
X.~Song, J.~Liu, Y.~Liu, Y.~Li, W.~Lei, R.~Wang, Rethinking radiology report generation via causal inspired counterfactual augmentation, in: Proceedings of the 15th ACM International Conference on Bioinformatics, Computational Biology and Health Informatics, 2024, pp. 1--10.

\bibitem{Camil}
K.~Chen, S.~Sun, J.~Zhao, Camil: Causal multiple instance learning for whole slide image classification, in: Proceedings of the AAAI Conference on Artificial Intelligence, Vol.~38, 2024, pp. 1120--1128.

\bibitem{CausalCLIPSeg}
Y.~Chen, M.~Wei, Z.~Zheng, J.~Hu, Y.~Shi, S.~Xiong, X.~X. Zhu, L.~Mou, Causalclipseg: Unlocking clip’s potential in referring medical image segmentation with causal intervention, in: International Conference on Medical Image Computing and Computer-Assisted Intervention, Springer, 2024, pp. 77--87.

\bibitem{wang2021bilateral}
C.~Wang, J.~Li, F.~Zhang, X.~Sun, H.~Dong, Y.~Yu, Bilateral asymmetry guided counterfactual generating network for mammogram classification, IEEE Transactions on Image Processing 30 (2021) 7980--7994.

\bibitem{richens2020improving}
J.~G. Richens, Improving the accuracy of medical diagnosis with causal machine learning, Nature communications 11~(1) (2020) 3923.

\bibitem{vae}
D.~P. Kingma, Auto-encoding variational bayes, arXiv preprint arXiv:1312.6114 (2013).

\bibitem{BUSI}
W.~Al-Dhabyani, M.~Gomaa, H.~Khaled, A.~Fahmy, Dataset of breast ultrasound images, Data in brief 28 (2020) 104863.

\bibitem{DDTI}
L.~Pedraza, C.~Vargas, F.~Narv{\'a}ez, O.~Dur{\'a}n, E.~Mu{\~n}oz, E.~Romero, An open access thyroid ultrasound image database, in: 10th International symposium on medical information processing and analysis, Vol. 9287, SPIE, 2015, pp. 188--193.

\bibitem{TUI}
Z.~Bai, L.~Chang, R.~Yu, X.~Li, X.~Wei, M.~Yu, Z.~Liu, J.~Gao, J.~Zhu, Thyroid nodules risk stratification through deep learning based on ultrasound images, Medical Physics 47~(12) (2020) 6355--6365.

\bibitem{ISIC2018}
M.~A. Al-Masni, D.-H. Kim, Multiple skin lesions diagnostics via integrated deep convolutional networks for segmentation and classification, Computer methods and programs in biomedicine 190 (2020) 105351.

\bibitem{KVASIR}
D.~Jha, P.~H. Smedsrud, M.~A. Riegler, P.~Halvorsen, T.~De~Lange, D.~Johansen, H.~D. Johansen, Kvasir-seg: A segmented polyp dataset, in: MultiMedia modeling: 26th international conference, MMM 2020, Daejeon, South Korea, January 5--8, 2020, proceedings, part II 26, Springer, 2020, pp. 451--462.

\bibitem{medt}
J.~M.~J. Valanarasu, P.~Oza, I.~Hacihaliloglu, V.~M. Patel, Medical transformer: Gated axial-attention for medical image segmentation, in: MICCAI 2021: 24th international conference, Springer, 2021, pp. 36--46.

\bibitem{ggnet}
C.~Xue, L.~Zhu, H.~Fu, X.~Hu, X.~Li, H.~Zhang, P.~A. Heng, Global guidance network for breast lesion segmentation in ultrasound images, Medical Image Analysis 70 (2021) 101989.

\bibitem{transunet}
J.~Chen, Y.~Lu, Q.~Yu, X.~Luo, E.~Adeli, Y.~Wang, L.~Lu, A.~L. Yuille, Y.~Zhou, Transunet: Transformers make strong encoders for medical image segmentation, arXiv preprint arXiv:2102.04306 (2021).

\bibitem{aaunet}
G.~Chen, L.~Li, Y.~Dai, J.~Zhang, M.~H. Yap, Aau-net: an adaptive attention u-net for breast lesions segmentation in ultrasound images, IEEE Transactions on Medical Imaging 42~(5) (2022) 1289--1300.

\bibitem{swinunet}
H.~Cao, Y.~Wang, J.~Chen, D.~Jiang, X.~Zhang, Q.~Tian, Swin-unet: Unet-like pure transformer for medical image segmentation, in: European conference on computer vision, Springer, 2022, pp. 205--218.

\bibitem{accunet}
N.~Ibtehaz, D.~Kihara, Acc-unet: A completely convolutional unet model for the 2020s, in: International Conference on Medical Image Computing and Computer-Assisted Intervention, Springer, 2023, pp. 692--702.

\bibitem{busseg}
H.~Wu, X.~Huang, X.~Guo, Z.~Wen, J.~Qin, Cross-image dependency modeling for breast ultrasound segmentation, IEEE Transactions on Medical Imaging 42~(06) (2023) 1619--1631.

\bibitem{EffiSegNet}
I.~Vezakis, K.~Georgas, D.~Fotiadis, G.~Matsopoulos, Effisegnet: Gastrointestinal polyp segmentation through a pre-trained efficientnet-based network with a simplified decoder, in: Proceedings of the International Conference of the IEEE Engineering in Medicine and Biology Society, IEEE, 2024, pp. 1--4.

\bibitem{LM-Net}
Z.~Lu, C.~She, W.~Wang, Q.~Huang, Lm-net: A light-weight and multi-scale network for medical image segmentation, Computers in Biology and Medicine 168 (2024) 1--12.

\bibitem{vgg16}
K.~Simonyan, Very deep convolutional networks for large-scale image recognition, arXiv preprint arXiv:1409.1556 (2014).

\bibitem{densenet}
G.~Huang, Z.~Liu, L.~Van Der~Maaten, K.~Q. Weinberger, Densely connected convolutional networks, in: Proceedings of the IEEE conference on CVPR, 2017, pp. 4700--4708.

\bibitem{resnet}
K.~He, X.~Zhang, S.~Ren, J.~Sun, Deep residual learning for image recognition, in: Proceedings of the IEEE conference on CVPR, 2016, pp. 770--778.

\end{thebibliography}



\end{document}